\documentclass[a4paper,12pt,titlepage]{article}
\usepackage{tabularx} 
\usepackage{amsmath}  
\usepackage{float} 
\usepackage{graphicx} 
\usepackage[margin=1in,letterpaper]{geometry} 
\usepackage{cite} 
\usepackage[final]{hyperref} 
\hypersetup{
	colorlinks=true,       
	linkcolor=blue,        
	citecolor=blue,        
	filecolor=magenta,     
	urlcolor=blue         
}

\usepackage[utf8]{inputenc}
\usepackage[inline]{enumitem}
\usepackage{subfigure}
\usepackage{comment}
\usepackage{longtable}

\newcommand{\degr}{^{\circ}}
\newcommand{\dd}{\textrm{d}}
\newcommand{\return}[0]{\vspace*{1em}}

\title{Detailed  Modelling  of  Ultraviolet  Radiation  (UV)  from  the  Interaction  of  Multiple  Lamps  in  Reactors, using Radiative Transfer Techniques}

\author{I. H. Cyr,  C.\ E.\ Jones \\ Department of Physics and Astronomy, The University of Western Ontario,\\ London, ON Canada N6A 3K7 \\ 
and \\
J. Robinson\\ Trojan Technologies, 3020 Gore Road,\\London ON Canada N5V 4T7}

\date{December 2020}
\begin{document}
\maketitle

\section{Introduction}

Trojan Technologies designs and manufactures UV-based water-treatment systems. The simplest systems contain a single UV lamp; more complex systems contain many lamps.
When there are multiple closely spaced lamps in a UV reactor, significant UV radiation emitted by the various lamps can reach, and “interact” with neighbouring lamps. Although many of the optical phenomena occurring in a UV reactor are well understood, and accounted for in Trojan’s current modelling, the fate of UV photons emitted from one lamp that reach the plasma of a neighbouring lamp after travelling through the intervening air, quartz, and water layers is still not well understood. These photons can be transmitted through the plasma, absorbed by the plasma and re-emitted, or absorbed by the plasma and not re-emitted. This work was undertaken with the goal of developing a model that accounts for these photon-plasma interactions to better predict the UV distribution throughout a reactor.

Some related experimental work has been done at Trojan in the past. This work is documented in Trojan internal reports RPT265 (2003), RPT393 (2006), RPT469 (2010), and TR0577 (2011). This earlier work provided some helpful insight, but was limited in scope and was subject to large experimental uncertainty. 

The work discussed herein began in December 2017 with NSERC Engage funding until May 31, 2018, and then continued under Mitacs Accelerate funding ending in January 2019. A presentation of the work to date was given at Trojan by Prof. Carol Jones on July 3, 2018 to a group of Trojan engineers and scientists. Early on in the work a literature review, including a review of Trojan’s previous work, was performed, and very helpful discussions were held with Prof. Georges Zissis of Paul Sabatier University, Toulouse, France. Prof. Zissis provided us with a simple computer code and valuable insight. We then developed a more-sophisticated Monte Carlo model; this model is described below. 

We continued the work under new Mitacs Accelerate funding beginning in February 2019. This work included making improvements to the model as well as performing comparisons of model predictions with actual UV measurements obtained in Trojan’s Lamp Lab for validation purposes. This work continued under a Mitacs Elevate award from February 2020 to January 2021. The model has now been satisfactorily validated, we plan to implement our results into software used by Trojan for their reactor modelling.

This report provides the background theory of radiative transfer in Section~\ref{theory} and includes a detailed description of the Monte Carlo code in Section~\ref{method}. A selection of relevant results is provided in Section~\ref{results} with a summary of our work and future directions presented in Sections~\ref{summary} and~\ref{future}, respectively. References are provided at the end of this document.

\section{Theory}
\label{theory}

Low-pressure mercury (Hg) lamps used by Trojan Technologies are tube-like devices of length ranging from a few cm to two meters, and diameters of approximately 16 to 38 mm. The interior of the lamp is filled with a mixture of Hg vapour and an inert gas such as argon. The vapour is contained in a cylindrical quartz wall approximately 1 mm thick. Electrodes are positioned at each end of the lamp, providing an arc current to activate the lamp.

Our current work is focused on 254-nm radiation in a low-pressure-Hg lamp. In these lamps, a 254-nm photon emitted from a Hg atom can be absorbed by another atom, exciting the absorbing atom\footnote{We assume here that, when a Hg atom absorbs a 254-nm photon, the atom was originally in the ground state.}. The excited atom can then emit another 254-nm photon, or be quenched (by non-radiative collisions), in which case the 254-nm photon is lost from the system. The initial model did not include an inert gas but the effects were indirectly included in later iterations of the model (see Section~\ref{sub:voigt}).

\subsection{Einstein coefficients}

Through the absorption or emission of a photon, atoms can transition from one energy state to another. Three main processes govern these transitions: spontaneous emission, absorption, and stimulated emission, each of which is associated with one of the three Einstein coefficients. Only the first two processes are relevant in this work.

Spontaneous emission is the process where an atom goes from a higher energy state (energy level 2) to a lower energy state (level 1) by emitting a photon, the energy of which will be equal to the energy difference of the two energy states, 

\begin{equation}
    \Delta E = E_2 - E_1 = h\nu_{21},
\label{eq:spont_em}
\end{equation}

\noindent where $\nu_{21}$ is the frequency of the emitted photon, and $h$ is Plank's constant.
\return 

The average decay time of an energy state, i.e. the probability of spontaneous emission per second, is given by the Einstein A-coefficient, $A_{ji}$, where $j$ and $i$ represent the upper and lower levels of the transition, respectively. The natural lifetime, $\tau$, of an energy level $j$ is related to the sum of the probabilities of spontaneous emission to all lower levels $i$,

\begin{equation}
    \tau_j = 1 \Big/ \sum_{i}{A_{ji}}.
\label{eq:tau}
\end{equation}

Absorption is the process in which an atom absorbs a photon and becomes excited into a higher energy state. For this process to occur, the energy of the photon must correspond to the difference in energy of the two states, similar to Equation~\ref{eq:spont_em}. The probability of absorption per second is given by $B_{12} I_{\nu}$, where $B_{12}$ is the Einstein B-coefficient of absorption and $I_{\nu}$ is the spectral intensity of the incident radiation in units of W s m$^{-2}$ sr$^{-1}$. 

The coefficients are related to each other by the following relation when the lower energy level (level 1) is the ground level,

\begin{equation}
    \frac{A_{21}}{B_{12}} = \frac{2 h \nu^3}{c^2} \frac{g_1}{g_2}
\end{equation}

\noindent where $c$ is the speed of light, and $g_1$ and $g_2$ are the statistical weights, also known as the quantum-mechanical degeneracies, of levels 1 and 2, respectively.

\subsection{Lineshape and line broadening}
\label{sub:lineshape}

The spectral lineshape describes how atoms, molecules, or ions of a certain medium affect the spectral intensity of incoming radiation. This lineshape is often described in terms of the absorption coefficient $\kappa$. 

This absorption coefficient can be related to the Einstein coefficients using the Ladenburg relation\cite{mol98},

\begin{equation}
    \int \kappa \left( \nu \right) \dd \nu = \frac{c^2}{8 \pi \nu_0^2} \frac{g_2}{g_1} N_1 A_{21},
\label{eq:ab_coef_int}
\end{equation}

\noindent where $\kappa \left( \nu \right)$ is the absorption coefficient as a function of frequency ($\nu$), $\nu_0$ is the frequency of the line centre, and  $N_1$ is the number density of lower-state atoms. Note that this version is a simplification of the Ladenburg relation, where we assume that the lower level is the ground level and that the number density of higher-state atoms ($N_2$) is very small compared to $N_1$, in other words in cases where stimulated emission is absent or negligible. In our case this assumption is valid.

A spectral line can be broadened by various mechanisms as discussed below. Absorption of a photon by an atom can occur if the frequency (wavelength) of the photon is within the frequency range of the broadened line. There are three major broadening mechanisms that typically contribute to the broadening of the spectral lineshape, depending on the parameters of the system under study: pressure broadening, Doppler broadening, and natural broadening.

For the Hg lamp in this work, the two most significant sources of broadening in the spectral lines are the Doppler and pressure broadening, the former being the largest of the two. Natural broadening is still present but its effects are too small comparatively and is therefore ignored to simplify the models.

\subsubsection{Doppler broadening}

Doppler broadening is due to the thermal motion of the gas atoms in the vapour. Because of this motion, a photon moving toward or away from a gas atom will appear to have its frequency shifted due to the Doppler effect if a component of its motion is parallel to the direction of the atom's motion.

The most probable velocity, $v_0$, of the atoms inside the vapour at a given (absolute) temperature, $T$, is given by,

\begin{equation}
    v_0 = \sqrt{\frac{2 k_B T}{m_{atom}}},
\label{eq:mav}
\end{equation}

where $k_B$ is the Boltzmann constant and $m_{atom}$ is the mass of a single atom. The lineshape $\kappa$ associated with the probability that a photon of frequency $\nu$ will be absorbed, due to Doppler broadening, is equal to

\begin{equation}
    \kappa \left( \nu \right) = \kappa_0^D \exp \left[- \frac{\left( \nu - \nu_0 \right)^2}{\nu_0^2} \frac{c^2}{v_0^2} \right],
\label{eq:k_dop}    
\end{equation}

where $\nu_0$ is the central frequency of the transition, and $\kappa_0^D$ is the Doppler contribution to the absorption coefficient at $\nu_0$.This equation is often written as \cite{mol98}

\begin{equation}
    \kappa \left( \nu \right) = \kappa_0^D \exp \left[- \left(\frac{2 \left( \nu - \nu_0\right)}{\Delta\nu^D} \sqrt{\ln{2}} \right)^2 \right],
\label{eq:k_dop_mod}    
\end{equation}

where 

\begin{equation}
    \Delta\nu^D = \frac{2}{c}\sqrt{\frac{2 \ln\left(2\right) k_B T}{m_{atom}}}\nu_0.
\label{eq:del_nu_d}    
\end{equation}

This form will become important later in the section. To determine $\kappa_0^D$, we substitute Equation~\ref{eq:k_dop} into Equation~\ref{eq:ab_coef_int} and integrate, 

\begin{equation}
    \kappa_0^D \left(\nu \right)  = \frac{c^3}{8 \pi^{3/2} \nu_0^3} \frac{g_2}{g_1} \frac{N_1}{\tau v_0}.
\end{equation}

Converting this previous equation in terms of wavelength, we are left with the following, 

\begin{equation}
    \kappa_0^D \left(\lambda \right) = \frac{\lambda_0^3}{8 \pi^{3/2}} \frac{g_2}{g_1} \frac{N_1}{\tau v_0}.
\end{equation}

Note that Equation~\ref{eq:k_dop} can also be converted to be in terms of wavelength, 

\begin{equation}
    \kappa \left( \lambda \right) = \kappa_0^D \exp \left[- \frac{\left( \lambda - \lambda_0 \right)^2}{\lambda^2} \frac{c^2}{v_0^2} \right].
\label{eq:k_dop_l1}    
\end{equation}

However, given that the difference between the wavelengths $\lambda$ and $\lambda_0$ is small,

\begin{equation}
   \frac{\left( \lambda - \lambda_0 \right)}{\lambda_0} \ll 1, 
\end{equation}

we can rewrite this Equation~\ref{eq:k_dop_l1} in similar form as Equation~\ref{eq:k_dop},

\begin{equation}
    \kappa \left( \lambda \right) \approx \kappa_0^D \exp \left[- \frac{\left( \lambda - \lambda_0 \right)^2}{\lambda_0^2} \frac{c^2}{v_0^2} \right].
\label{eq:k_dop_l}    
\end{equation}

\subsubsection{Pressure broadening}

Pressure broadening is due to the disturbance of energy levels caused by collisions with other atoms. In our case these collisions can occur with either other Hg atoms or with atoms of the filling gas\footnote{We note that the presence of a filling gas is not directly taken into account in this model, but its effects are accounted for indirectly in the Voigt parameter (discussed below).}.

There are two methods to describe this type of broadening, the statistical approach, as described by Berman and Lamb (1969) \cite{ber69}, and the collisional (Lorentz) theory. The latter is a simplification of the statistical approach which works best in low pressure conditions like those found inside low-pressure Hg lamps, and is therefore the method used in this work.

The absorption function associated with this type of broadening can be expressed using a Lorentz theory \cite{mol98},

\begin{equation}
    \kappa \left( \nu \right) = \frac{\kappa_0^{L}}{1+ \left[ \frac{2 \left(\nu - \nu_0 \right)}{\Delta \nu^{coll}}\right]^2}, 
\label{eq:k_Lorentz}    
\end{equation}

where $\kappa_0^L$ is the Lorentz contribution to the absorption coefficient at $\nu_0$, and $\Delta \nu^{coll}$ is the full width at half maximum (FWHM) of the Lorentzian distribution, the latter being proportional to the species collision rate ($Z_L$) \cite{and85};

\begin{equation}
    \Delta \nu^{coll} = \frac{Z_L}{\pi}. 
\label{eq:delta_nu_L}    
\end{equation}

\subsubsection{Voigt profile}
\label{sub:voigt}

Doppler and pressure broadening use two different kinds of distribution, that is a Gaussian (or normal) distribution and a Lorentzian distribution, respectively. Therefore the combination of both types of broadening can be described by the convolution of both distributions. The convolution of a Gaussian function with a Lorentzian function is called a Voigt profile. We will not go into detail here on how this function is derived (see Mitchell and Zemansky 1961, p.100-101 \cite{mit61} for more detail), but the Voigt profile can be written as

\begin{equation}
    \kappa \left( \nu \right) = \kappa_0^{D} \frac{a}{\pi} \int_{-\infty}^{\infty} \frac{\exp \left( -y^2\right)}{a^2 + \left[ \frac{2\left(\nu - \nu_0\right)}{\Delta \nu^D} \sqrt{\ln{2}} - y \right]} \mathrm{d}y,
\label{eq:k_voigt}    
\end{equation}

where $a$ is the Voigt parameter\footnote{This parameter takes the collision rate into account and therefore indirectly accounts for the effect of the filling gas.}, which indicates in this case the relative importance of the Doppler and pressure broadening,

\begin{equation}
    a = \sqrt{\ln{2}} \frac{\Delta \nu^{coll}}{\Delta \nu^D}.
\label{eq:voigt_param}    
\end{equation}

When using the Voigt profile in this work, we have adopted a Voigt parameter value of 0.02 as prescribed in Anderson 1985 \cite{and85}.

The integral in Eq~\ref{eq:k_voigt} cannot easily be reduced to a simple numerical formula. However it can be approximated using the following equations \cite{and85}.

\begin{equation}
    \kappa \left( \lambda \right) \approx \kappa_0^D \frac{\left[ V\left( Z\right) / C \right]_{a=a, Z=Z}}{\left[ V\left( Z\right) / C\right]_{a=0, Z=0}},
\label{eq:voigt_approx}    
\end{equation}
where

\begin{center}
\begin{equation}
 \begin{split}
    C &= \left[ 2.12839 \exp\left(- a / 1.23 \right) + \pi \left(1 - \exp\left((- a / 1.23  \right) \right) \right] W,\\
    V\left( Z\right) &= \exp\left(- a / 1.23 \right) D\left( Z\right) + \left(1 - \exp\left((- a / 1.23  \right) \right) L\left( Z\right),\\
     L\left( Z\right) &= \frac{1}{1+Z^2},\\
     D\left( Z\right) &= \exp \left( - 0.69315 Z^2\right),\\
     Z &= \frac{X}{W},\\
     X &= \frac{c}{v_0} \left( \frac{\lambda_0}{\lambda} - 1\right),\\
     W &= \frac{a + a^2}{2+a} + 0.83255.
\label{eq:approx}    
\end{split}
\end{equation}
\end{center}

Note that this approximation only holds for $a > 0.01$.

\subsection{Mercury vapour}

\begin{figure}[!htbp]
\centering
\includegraphics[width=0.8\textwidth]{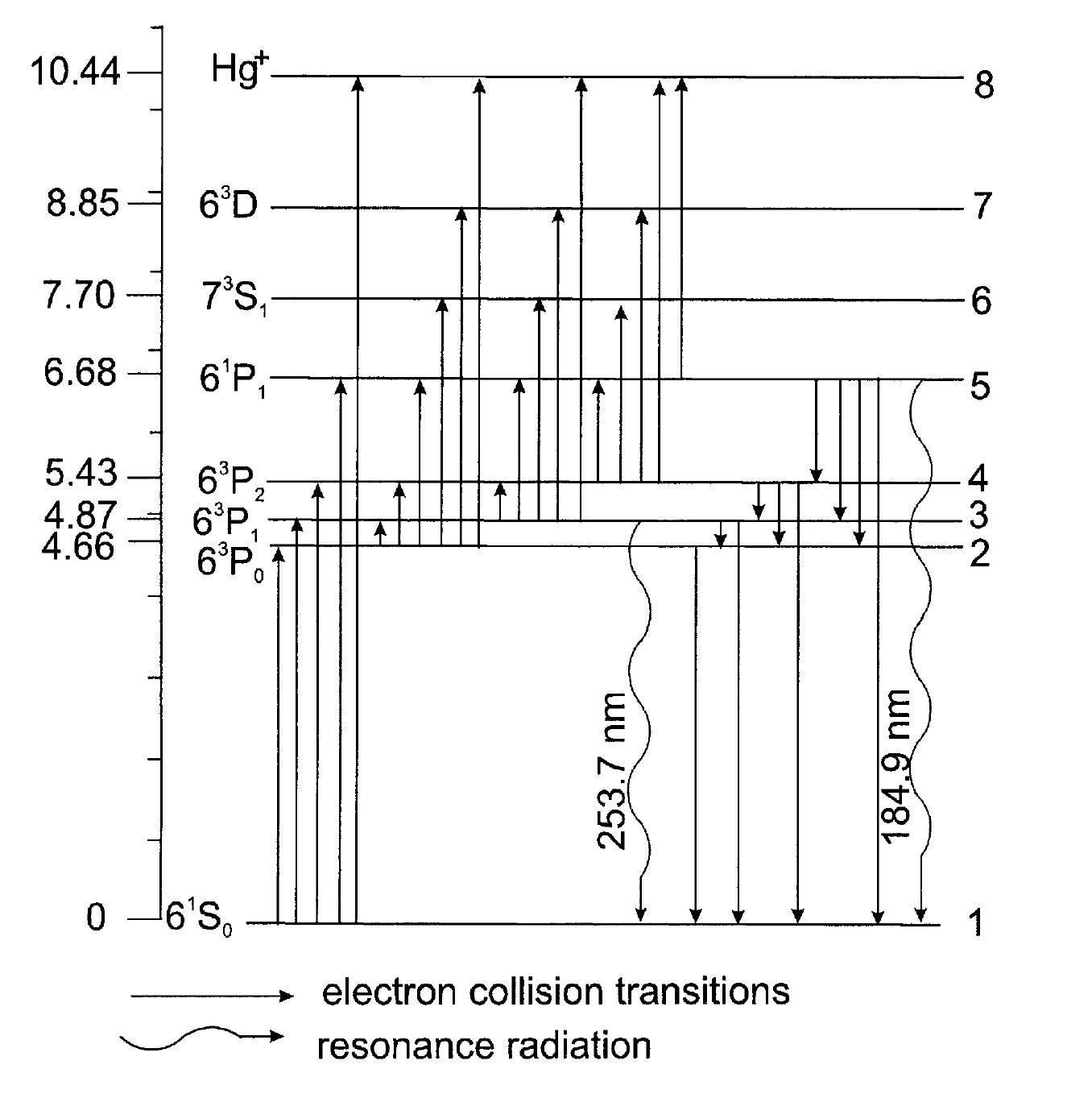}
\caption{Energy level diagram for Hg atoms, taken from Baeva and Reiter (2003)\cite{bae03}. The left axis shows the energy, in electron Volts, of each energy level relative to the ground level.}
\label{fig:energylevel}
\end{figure}

As the name implies, Hg is the main component of low-pressure Hg lamps. Figure~\ref{fig:energylevel}, taken from Baeva and Reiter (2003)\cite{bae03}, shows a diagram of the various energy levels of a Hg atom and possible transitions between these levels. For these transitions, only a few result in the generation of photons. The most important for this work, is the 253.7 nm transition, i.e. the transition from energy level 6$^3$P$_1$ to 6$^1$S$_0$, as seen in Figure~\ref{fig:energylevel}.

In nature Hg comes in seven different stable isotopes, each one of which has a slightly different central wavelength. Furthermore, two of these isotopes have an odd mass number, which causes these isotopes to have more than one state with different central wavelength due to nuclear magnetic splitting. Therefore we count all of these as separate species of Hg. Table~\ref{table:isotope} shows the abundances and central wavelength of the 6$^3$P$_1$ $\rightarrow$ 6$^1$S$_0$ transition for each species of Hg. For this work, we only consider the absorption from non-excited Hg atoms.

\begin{table}[!htbp]
\begin{center}
\caption{Stable Hg species}
\begin{tabular}{ccc} 
\hline \hline \\[-2ex]
Mass numbers & Abundance & Central wavelength \\
(amu) & Refs. \cite{bae03} \cite{vei83} \cite{hay12}& (nm)\\[0.5ex]
\hline \\[-1.8ex]
	196 & 0.0015 & 253.6508\\
	198 & 0.0997 & 253.6517\\
	199a & 0.0562 & 253.6550\\
	199b & 0.1125 & 253.6503\\
	200 & 0.2310 & 253.6527\\
	201a & 0.0659 & 253.6548\\
	201b & 0.0439 & 253.6518\\
	201c & 0.0220 & 253.6502\\
	202 & 0.2986 & 253.6539\\
	204 & 0.0687 & 253.6550\\
\\[-1.8ex] \hline \hline  	

\end{tabular}
\label{table:isotope}
\end{center}
\end{table}

In a (non-amalgam) low-pressure Hg lamp, the number density and pressure of Hg vapour are determined by the coldest spot on the lamp wall, known as the cold-spot. In the case of an amalgam lamp, the pressure is determined by the temperature and makeup of the amalgam. The relation between these quantities is rather complex. For this work, we used values calculated in the literature\cite{hub06} for the non-amalgam lamp, and values taken from amalgam phase diagrams for the amalgam lamps.


\section{Methodology}
\label{method}

The Monte Carlo method is a stochastic method that easily allows for a flexible geometry and works particularly well on statistical problems. We adopted this method to tackle this problem since the size and distribution of lamps within reactors can be changed to mimic any particular arrangement of interest. Our Monte Carlo code generates millions of photons and follows their absorption and re-emission throughout the medium of the lamp so that we can predict the distribution of resultant photons. 

Two cases are modelled: active lamp and inactive lamp. In the active-lamp case, photons are generated inside the lamp, while in the inactive-lamp case, the lamp does not generate original photons, but is irradiated with photons generated by an external source. In addition, photons generated externally can be used to irradiate an active lamp, and the source of externally generated photons (in either the active- or inactive- lamp case) can be an active lamp.

The methodology for an inactive lamp is identical to that for an active lamp except that for an inactive lamp, the initial photon properties are determined by the properties of the external source of the photons, and there are no excited Hg atoms prior to exposure to the externally generated photons. In both cases, the distribution of unexcited Hg atoms is uniform.

\subsection{Photon generation (active-lamp case)}

During the generation of photons, their properties are generated randomly, selected from a certain distribution related to the physical conditions of the current simulation. These properties are the photon's initial position, its direction of motion, and its wavelength or frequency.

\subsubsection{Initial position}

The position of a photon can be denoted using cylindrical ($r,\phi,z$) and Cartesian ($x,y,z$) coordinates. We begin with a cylindrical system to generate the initial position of each new photon and then convert to Cartesian coordinates for convenience.

The initial radial position, $r_0$, is determined by the distribution of excited-state Hg atoms in the gas. To describe the radial distribution of excited-state Hg atoms, a Bessel function of the first kind, $J_0 \left( 2.4 r/R \right)$, is used, where $R$ is the radius of the lamp\footnote{This function was suggested by Dr. Georges Zissis in private communications.}. In order to generate $r_0$ which follows the same distribution, we first need to determine the inverse cumulative distribution function (or quantile function) associated with this distribution. This quantile function, $Q_{J_0}$, is determined through curve fitting, and is best approximated by the following function:

\begin{equation}
    Q_{J_0} \left(u \right) = 1 - \frac{2}{\pi} \arccos(u),
\end{equation}

\noindent where $u$ is a random variable with a uniform distribution between 0 and 1. The initial radial position of the photon can therefore be generated using the following expression,

\begin{equation}
    r_0 = Q_{J_0} \left(u \right) R,
\end{equation}

\noindent where $u$ is again a randomly generated number between 0 and 1. 
\return

The initial azimuthal and longitudinal positions of the photon, $\phi_0$ and $z_0$, respectively, are generated similarly since our model assumes a uniform distribution in excited-state Hg atoms along these two components. Therefore all we need to do is generate a uniformly generated random number scaled to the range of each component: 0 to 2$\pi$ for $\phi_0$ which is given by,
\begin{equation}
    \phi_0 = 2 \pi u,
\end{equation}
and $-L_z/2$ to $L_z/2$ for $z_0$ given by
\begin{equation}
    z_0 = L_z \left( w - 1/2 \right),
\end{equation}
\noindent where $L_z$ is the length of the lamp, and $u$ and $w$ are independently generated random numbers between 0 and 1. 

Once we have the position of the photon in cylindrical coordinates ($r,\phi,z$), we can easily calculate its position in Cartesian coordinates ($x, y, z$), which will be more convenient when calculating the propagation of the photon between each re-absorption event,

\begin{equation}
\begin{split}
x &= r \cos \left( \phi \right), \\
y &= r \sin \left( \phi \right). \\
\end{split}
\end{equation}

\noindent Note that the $z$ component is the same in both systems.

\subsubsection{Initial direction of travel}
\label{sub:dir_angle}

The direction of travel of a photon in a three dimensional space can be described using two angular components: an azimuthal direction\footnote{Not to be confused with the azimuthal position.}, $\mu$, which describes the direction of travel perpendicular to the length of the lamp with a range of [0,2$\pi$], and a polar direction, $\omega$, describing the direction of travel in the longitudinal plane with range [$0,\pi$]. 

\begin{figure}[!htbp]
\centering
\includegraphics[width=0.8\textwidth]{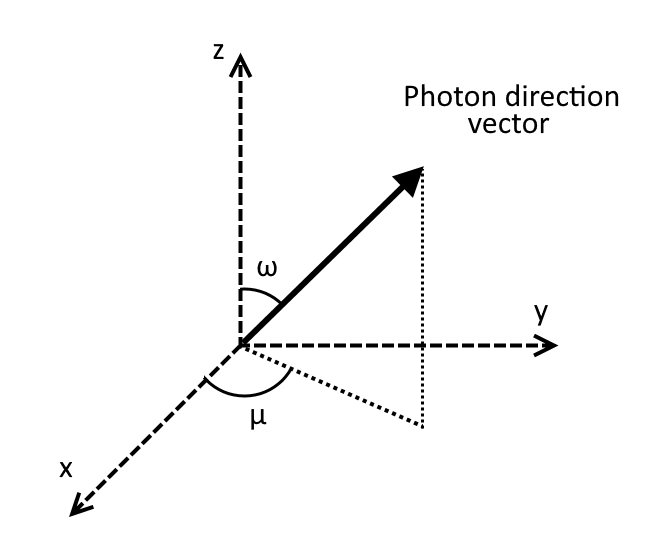}
\caption{Diagram representing the two direction angles $\mu$ and $\omega$.}
\label{fig:dirvec}
\end{figure}

The method of picking a random direction is similar to picking a random point on the surface of a sphere. In order to ensure an even distribution over the area of the sphere, we must adopt an $\arccos$ distribution for $\omega$, favouring the equatorial area over the poles, since using a uniform distribution would result in an oversampling of the poles.  

\begin{equation}
    \mu = 2 \pi u,
\end{equation}
\begin{equation}
    \omega =  \cos^{-1}\left( 2 w - 1 \right),
\end{equation}

\noindent where $u$ and $w$ are random numbers between 0 and 1.

\subsubsection{Initial wavelength}
\label{sub:init_wl}

The generation of the initial wavelength of a photon is accomplished in two steps. First we determine from which isotope the photon is emitted. The list of isotopes along with their typical abundance ratios are presented in Table~\ref{table:isotope}. We use these abundances to weight the probability that the photon is emitted by that particular isotope. To randomly select an isotope, we generate a uniform random number, $u$, between 0 and 1. To determine which isotope corresponds to $u$, we first compare it to the probability of the first isotope, $P^{iso}_1$. If the number generated is lower than $P^{iso}_1$, the first isotope is selected. If the number generated is higher than $P^{iso}_1$, this probability is subtracted from the number generated and the result is compared to the next probability value. This process is repeated until an isotope is selected.

Once the emitting isotope is chosen, a wavelength value is randomly selected based on the lineshape equation relevant for the desired model, either Eq~\ref{eq:k_dop} for a purely Gaussian profile (Doppler broadening only) or Eq~\ref{eq:k_voigt} for a Voigt profile (Doppler and pressure broadening). Because each isotope has a different central wavelength, as seen in Table~\ref{table:isotope}, the lineshape is different for each isotope. 

In order to generate a random number which follows a Gaussian distribution, $\lambda$, we first generate two random numbers, $u$ and $v$, taken from a uniform distribution between 0 and 1. Then we generate our desired Gaussian distributed random number using the Box–Muller transform\cite{box58},

\begin{equation}
    \lambda =  \sigma_i \sqrt{-2.0  \ln \left(1.0 - u \right)} \cos\left(2 \pi v \right) + \lambda_{i,0},
\end{equation}

where $i$ represents a specific isotope, $\lambda_{i,0}$ is the central wavelength related to isotope, $i$, and $\sigma_i$ is related to the width of the lineshape for isotope $i$. The latter can be determined using Equation~\ref{eq:k_dop_l}, which is the form of a Gaussian,

\begin{equation}
    \sigma_i =  \frac{\lambda_{i,0} v_0}{ \sqrt{2} c}.
\end{equation}

A similar method is used to generate a random number following the Voigt distribution profile. This time we require three random variables, $u$, $v$, and $w$, again uniformly distributed between 0 and 1. Using the Box–Muller transform as a base, Lee 1974 \cite{lee74} derived the following equation to generate a random variable, $x$, which follows the Voigt distribution, 

\begin{equation}
    x =  a \tan\left[ \left( w - 0.5\right) \pi\right] +  \sqrt{-\ln \left(u \right)} \cos\left(2 \pi v \right).
\end{equation}

We can then transform this variable $x$ into a wavelength $\lambda$ by using this formula, 

\begin{equation}
    \lambda = \frac{c \lambda_{i,0}}{x v_0 + c}.
\end{equation}

\subsection{Photon path}
\label{sub:photon_path}

Once the photon is generated, or in the inactive lamp case when the photon enters the lamp, it will travel through the lamp until it either is absorbed by another Hg atom or escapes the lamp. The free flight length, $l$, of a photon before being absorbed is determined by the following equation,

\begin{equation}
    l =  - \frac{\ln w}{\sum_i \kappa_i \left( \lambda \right)},
\end{equation}

where $w$ is a uniform random number between 0 and 1, and $\kappa_i\left( \lambda \right)$ is the absorption coefficient function for isotope $i$. Therefore, the sum represents the total absorption coefficient for all isotopes at the wavelength of the photon, $\lambda$.

In order to determine which isotope absorbed the photon, we use the same numerical technique described in Section~\ref{sub:init_wl} to select the initial isotope, but instead of weighting isotopes by abundances, $P^{iso}_i$, we weight them by their contribution to the absorption coefficient, $\kappa_i$.

Using the initial position of the photon ($x_i,y_i,z_i$) and its direction ($\mu, \omega)$, we can calculate its final position ($x_f,y_f,z_f$) after traveling a distance of $l$ through the lamp,

\begin{equation}
\begin{split}
x_f &= x_i + l \cos \left( \mu \right) \sin \left( \omega \right),\\
y_f &= y_i + l \sin \left( \mu \right) \sin \left( \omega \right),\\
z_f &= z_i + l \cos \left( \omega \right),\\
\end{split}
\label{eq:final_pos}
\end{equation}

which can easily be converted into cylindrical coordinates using,

\begin{equation}
\begin{split}
r_f &= \sqrt{x_f^2 + y_f^2},\\
\phi_f &= \tan^{-1}\left( \frac{y_f}{x_f} \right). \\
\end{split}
\label{eq:final_r_phi}
\end{equation}

If the final position of the photon is inside the radius and length of the lamp, the photon is reabsorbed by another Hg atom (see Section~\ref{sub:abs_reem}); if not, the photon reaches the lamp wall (see Section~\ref{sub:wall}).

\subsection{Absorption and re-emission}
\label{sub:abs_reem}

Once absorbed, we assume one of three things will happen to the photon: 1) it will be re-emitted by the same atom, 2) its energy will be transferred to another atom before being re-emitted, or 3) it will be quenched, meaning its energy is lost due to some non-radiative process. We denote the probability of a photon being re-emitted, transferred, or quenched with $P_{em}$, $P_{tr}$, and $P_{qu}$, respectively, with each being a number between 0 and 1, with their sum equal to 1. The values used in this work are taken from the Anderson et al. (1985)\cite{and85} and given in Table~\ref{table:prob}.

\begin{center}
\begin{longtable}{lcc}
\caption{Probability values for Hg vapour}  
\label{table:prob}\\
\hline \hline \\[-2ex]
 & \multicolumn{2}{c}{Wall temperature}\\
Probabilities & 15$\degr$C & 40$\degr$C \\[0.5ex]
\hline \\[-1.8ex]

Re-emission & 0.9901 & 0.9463\\

Transfer & 0.0060 & 0.0500\\

Quenching & 0.0039 & 0.0037\\

\\[-1.8ex] \hline \hline  	
\end{longtable}
\end{center}

In the first two cases, the photon will be re-emitted. Its new initial position is set to the final position calculated using Equations~\ref{eq:final_pos} and \ref{eq:final_r_phi}, while its new set of direction angles, $\mu$ and $\omega$ are generated in the same way as described in Section~\ref{sub:dir_angle}. Its wavelength is generated as described in Section~\ref{sub:init_wl}, but the isotope depends on whether the energy of the photon was transferred to another isotope (case 2), and in this case a new isotope is randomly selected as described in Section~\ref{sub:init_wl}, or if not (case 1), the absorbing isotope is determined as described in Section~\ref{sub:photon_path}.

\subsection{Wall interaction}
\label{sub:wall}

The gas interior of a low-pressure Hg lamp is enclosed within the lamp wall. In UV lamps, the wall is typically a quartz tube with circular cross section. Because quartz has a different refractive index than the interior and exterior of the lamp ($n_{quartz} = 1.505$), some photons will be reflected when they are incident on either the inner or outer surface of the lamp wall. Photons that are not reflected will be refracted and transmitted (or simply transmitted if their direction is normal to the interface).

\begin{figure}[!htbp]
\centering
\includegraphics[width=0.6\textwidth]{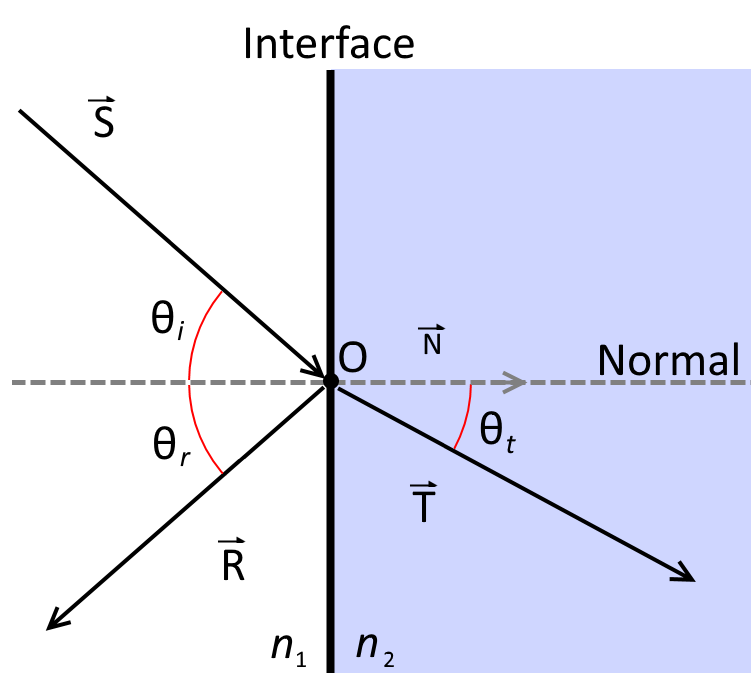}
\caption{Diagram of the interaction of incident radiation, $\vec{S}$, with the interface of two media of refractive indices $n_1$ and $n_2$, showing the part of the radiation that is transmitted, $\vec{T}$, and the part reflected, $\vec{R}$. The vector $\vec{N}$ represents the vector normal to the interface.}
\label{fig:fresnel}
\end{figure}

Figure~\ref{fig:fresnel} shows what happens when incident radiation, $\vec{S}$, hits the interface between two different media with different refractive indices, $n_1$ and $n_2$, at an angle $\theta_i$ from the normal of the interface. The vectors $\vec{R}$ and $\vec{T}$ represent the paths of the reflected and transmitted parts, respectively, of the incident radiation after coming into contact with the interface. According to the law of reflection\cite{hec02},

\begin{equation}
    \theta_r = \theta_i,
\end{equation}

\noindent while the relation between $\theta_t$ and $\theta_i$ can be determined using Snell's law,

\begin{equation}
n_1 \sin \theta_i = n_2 \sin \theta_t.
\label{eq:snell}
\end{equation}

The probability of an incident photon being reflected or transmitted through the interface can be determined using the Fresnel equations. These equations consider two cases, that is, whether the electric field of the incident radiation has a polarization parallel or perpendicular, denoted by $p$ and $s$ respectively, to the plane of incidence. The equation for the reflectance for each of these cases is, 

\begin{equation}
\begin{split}
    R_p &= \left| \frac{n_1 \cos \theta_t - n_2 \cos \theta_i}{n_1 \cos \theta_t + n_2 \cos \theta_i} \right|^2\\
    R_s &= \left| \frac{n_1 \cos \theta_i - n_2 \cos \theta_t}{n_1 \cos \theta_i + n_2 \cos \theta_t} \right|^2.
\end{split}
\label{eq:reflection_ps}
\end{equation}

For our purposes, it is assumed that the photons do not have a preferred polarization. Therefore the effective reflectance, $R_{eff}$, is given by,

\begin{equation}
R_{eff} = \frac{R_p + R_s}{2}.
\label{eq:ref_eff}
\end{equation}

This reflectance value gives us the probability that a photon will be reflected at the interface. Because it is assumed that no photons are lost at the interface, the probability of a photon passing through the interface (transmittance) is given by,

\begin{equation}
T_{eff} = 1 - R_{eff}.
\label{eq:tr_eff}
\end{equation}

Because the walls of the lamp are curved in the $x-y$ plane, the direction of the vector normal to the interface, $\vec{N}$, will be dependant on the azimuthal position $\phi$ at which the photon hits the wall. This vector can be written as follows,

\begin{equation}
\hat{N} \equiv 
\begin{pmatrix}
  N_x \\
  N_y \\
  N_z
 \end{pmatrix}
=
 \begin{pmatrix}
  -\cos(\phi) \\
  -\sin(\phi) \\
  0
 \end{pmatrix}.
\end{equation}

Similarly, we can write the direction vector of the incident photon using both direction angles $\mu$ and $\omega$,

\begin{equation}
\hat{S} = 
 \begin{pmatrix}
  \cos(\mu) \sin(\omega) \\
  \sin(\mu) \sin(\omega) \\
  \cos(\omega)
 \end{pmatrix}.
\end{equation}

Note that both vectors are unit vectors and the coordinates are centred at the point where the photon comes into contact with the interface (point $O$ in Figure~\ref{fig:fresnel}).

The angle between the incident photon and the normal, $\theta_i$, can be determined by taking the vector (cross) product of the two vectors,

\begin{equation}
\sin \left( \theta_i \right) = \frac{\left| \hat{N} \times \hat{S} \right|}{\left|\hat{N} \right| \left| \hat{S}\right|}.  
\end{equation}

If the photon is reflected back into the lamp, we use a basic rotation matrix $\mathbf{M}_r$ to determine the new direction of the photon, 

\begin{equation}
\hat{R} = \mathbf{M}_{r} \hat{S},
\label{eq:rotation}
\end{equation}

where

\begin{equation}
\mathbf{M}_{r} = \begin{bmatrix}
 2 \sin^2\phi - 1 & -2 \sin\phi \cos\phi  & 0 \\ 
 2 \sin\phi \cos\phi &  2 \cos^2\phi & 0\\ 
 0 & 0 & 1
\end{bmatrix}.
\label{eq:rot_matrix}
\end{equation}

The new set of direction angles is then calculated from the resulting vector and the radiative transfer processes are continued (see Section~\ref{sub:photon_path}).

If the photon is transmitted through the interface and into the quartz wall, its new trajectory vector is calculated using the following equation, which is based on Snell's law\cite{mik12},

\begin{equation}
\hat{T} = \frac{n_1}{n_2} \left[ \Hat{N} \times \left( - \Hat{N} \times \Hat{S} \right) \right] - \hat{N} \sqrt{1 - \left( \frac{n_1}{n_2} \right)^2 \left( \Hat{N} \times \Hat{S}\right) \cdot \left( \Hat{N} \times \Hat{S}\right)}.
\end{equation}

Once the photon is inside the quartz wall, it will travel in a straight line until either it hits the outer interface or it is absorbed by the quartz. Each time the photon traverses the quartz from the inner to the outer interface, or vice versa, it has a chance of being absorbed by the quartz. Every time a photon traverses a distance, $l$, inside the quartz, its probability of being absorbed, $P_{abs}$, is given by,

\begin{equation}
P_{abs} = 1 - 10^{-l Q_{abs}},
\label{eq:q_abs}
\end{equation}

where $Q_{abs}$ is the absorption coefficient of quartz at approximately 253.65 nm. The reflection and refraction processes are repeated for the outer interface of the wall. If the photon is transmitted through this interface, it is considered to have left the lamp. If the photon is reflected, the reflection, refraction, and absorption processes are repeated until the photon either exits the lamp or re-enters the lamp interior.

\section{Results}
\label{results}

In this section, we present relevant results obtained during the various stages of development of the code. We started with a very basic lamp model and tested different elements and scenarios with the model. Once we were reasonably certain that these elements were implemented correctly, more complex elements were added to the model and testing continued. For example, our initial lamp model consisted simply of a single isotope of Hg inside a circular enclosure, and only the position of the photons along the $r-\phi$ plane was considered, i.e. initially the longitudinal motion of photons was not tracked. Also, the walls of the lamp were initially assumed to be perfectly transparent prior to implementing a more realistic quartz wall. 

\subsection{Preliminary testing of the Monte Carlo code}
\label{sub:pretest}

The goal of the tests described below was to verify that the basic radiative transfer parts of the code were working as expected, with a focus on absorption and re-emission of photons. For these preliminary tests, we simulated the emission of photons from within a single lamp (i.e. there was no secondary source of photons). Photons were emitted from inside the lamp from random isotopes (weighted by abundance) at random wavelengths and these were assigned random directions (see Section~\ref{method}). This test will be hereinafter referred to as the single-lamp test.

\begin{figure}[!htbp]
\centering
\includegraphics[width=0.8\textwidth]{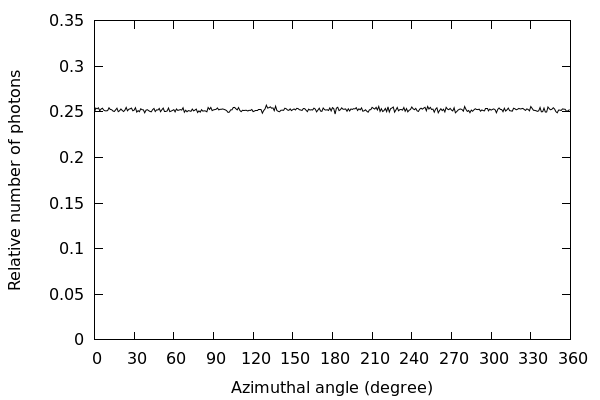}
\caption{Relative number of photons exiting the lamp as a function of azimuthal angle in a single-lamp test.}
\label{fig:phi_sgl}
\end{figure}

\begin{figure}[!htbp]
\centering
\includegraphics[width=0.8\textwidth]{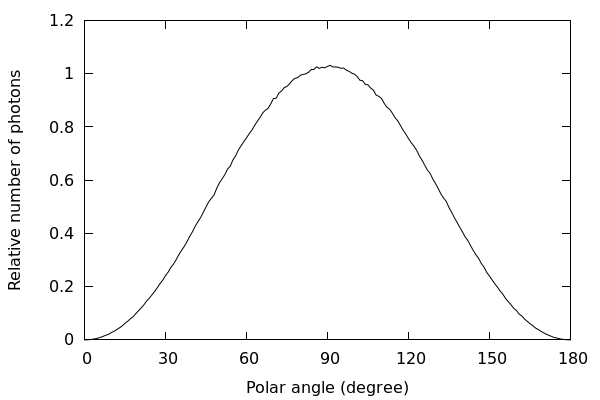}
\caption{Relative number of photons exiting the lamp as a function of polar angle in a single-lamp test.}
\label{fig:theta_sgl}
\end{figure}

Figure~\ref{fig:phi_sgl} shows the relative number of photons exiting the lamp as a function of their azimuthal angle $\phi$ for a single-lamp test, while Figure~\ref{fig:theta_sgl} shows the relative number of photons exiting the lamp as a function of their polar angle $\theta$ for the same test.

In Figure~\ref{fig:phi_sgl} we see that the relative number of photons is approximately the same in all directions, which is what we expect since the lamp is assumed to be axisymmetric. On the other hand, Figure~\ref{fig:theta_sgl} shows a variation in the polar angle of escaping photons, with the majority centered at 90$\degr$ (perpendicular to the lamp's length), and dropping completely near 0$\degr$ and 180$\degr$ (parallel to the lamp's length). This is also consistent with our expectations. The geometry of the lamp is such that the shortest path between the photon and the wall, i.e. the path with the lowest probability of re-absorption, is always at a polar angle of 90$\degr$. Combining this with the fact that the direction of the photons is reset after every absorption-reemission event means that photons have a higher likelihood of escaping the confinement of the lamp when their polar direction is near 90$\degr$.

The second preliminary test (simulation) consisted of an inactive lamp being irradiated from an outside source. Recall that an inactive lamp is defined as an unpowered lamp that does not generate photons on its own.

\begin{figure}[!htbp]
\centering
\includegraphics[width=0.8\textwidth]{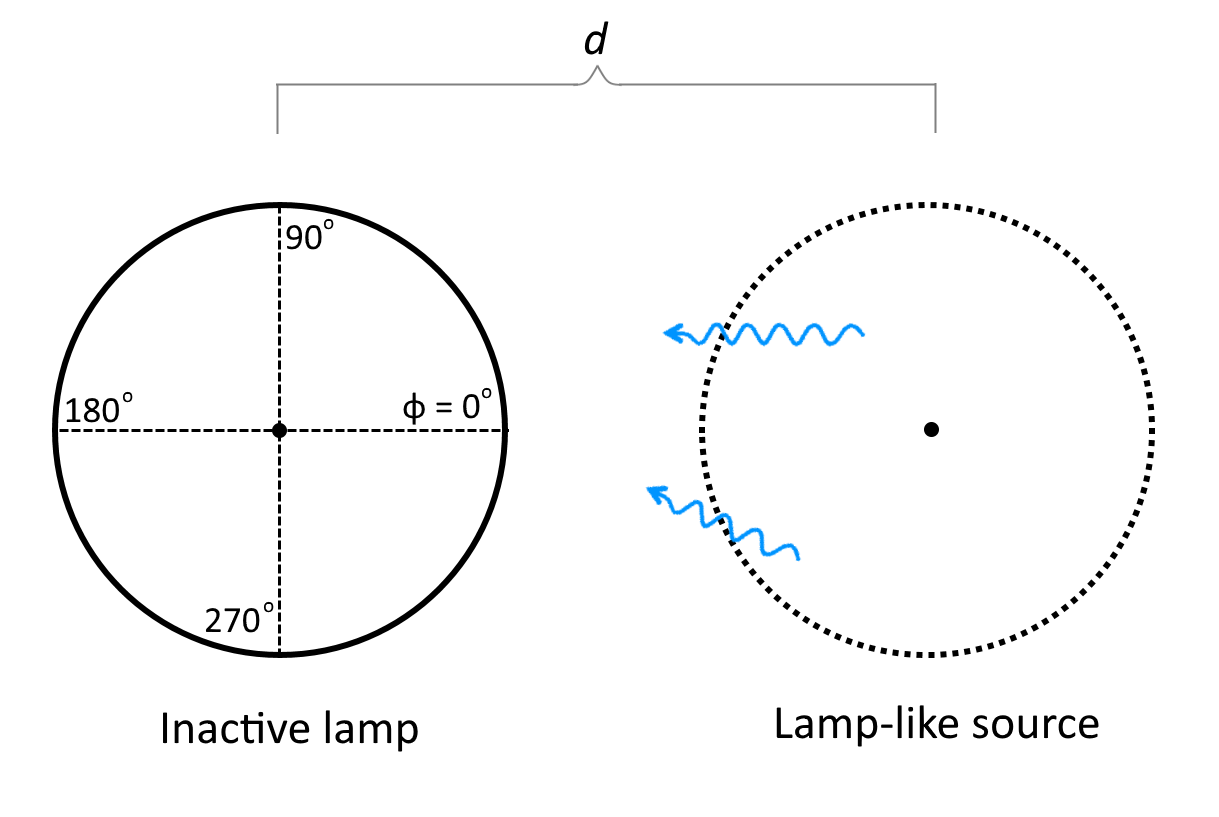}
\caption{Cross section view of an inactive lamp and a lamp-like source used in preliminary tests.}
\label{fig:ll_test}
\end{figure}

During this testing we looked at various types of outside sources, such as a single point source, a line of point sources, or a volumetric distributed source. The lamp-like source is placed parallel to the inactive lamp, at a distance $d$ from it (see Figure~\ref{fig:ll_test}). Photons are generated inside the source just as they would in an active lamp; however, it is assumed for simplicity that they do not interact with the medium of the lamp-like source. This is of course not representative  of real experimental scenarios; however, it provides insight by allowing us to follow the photons that interact with the inactive lamp. Furthermore, because not every photon coming out of the source can hit the inactive lamp, we adjusted the number of photons emitted by the source in order to have as many photons hitting the inactive lamp as there were photons generated in the active lamp. This number of photon remains constant no matter the active lamp diameter or its distance to the inactive lamp.

\begin{figure}[!htbp]
\centering
\includegraphics[width=0.8\textwidth]{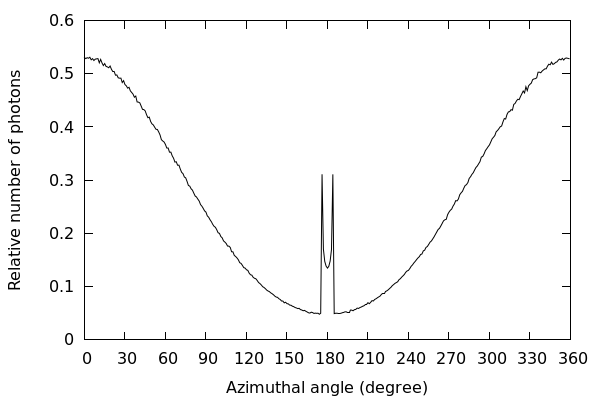}
\caption{Same as Figure~\ref{fig:phi_sgl} but for a two-lamp test.}
\label{fig:phi_sec}
\end{figure}

\begin{figure}[!htbp]
\centering
\includegraphics[width=0.8\textwidth]{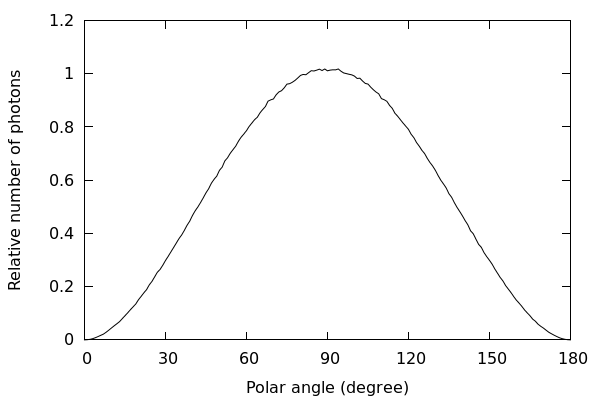}
\caption{Same as Figure~\ref{fig:theta_sgl} but for a two-lamp test.}
\label{fig:theta_sec}
\end{figure}

Similar to Figures~\ref{fig:phi_sgl} and \ref{fig:theta_sgl},  Figures~\ref{fig:phi_sec} and \ref{fig:theta_sec} show the number of photons exiting the inactive lamp as a function of their azimuthal ($\phi$) and polar ($\theta$) angles, respectively. The lamp-like source used to obtain the results shown in Figures~\ref{fig:phi_sec} and \ref{fig:theta_sec} was a volumetric distributed source.

In contrast to Figure~\ref{fig:phi_sgl}, Figure~\ref{fig:phi_sec} shows a much bigger dependence on the azimuthal direction of exiting photons, with the greatest concentration centred around 0$\degr$ and dropping significantly towards 180$\degr$. This is due to the fact that incident photons come from the $\phi = 0\degr$ side, and therefore are more likely to be absorbed and re-emitted near that side of the lamp, and consequently are more likely to exit on that side. The feature in the region of 180$\degr$ is caused by photons that pass through the lamp without being absorbed. The width and height of this feature is dependant on the diameter of the lamp-like source as well as its distance from the inactive lamp, with the width increasing and height decreasing with increasing diameter and/or decrease in distance, and vice versa. These width and height variations correlate with the angle at the incidence point (on the inactive lamp) subtended by the cross-section of the lamp-like source; they are not unexpected, keeping in mind that the range of incidence angles increases with increasing subtended angle, and that the total number of incident photons is the same in all cases. Figure~\ref{fig:theta_sec} on the other hand is very similar to its counterpart, Figure~\ref{fig:theta_sgl}. 

Note that Figures~\ref{fig:phi_sgl}, \ref{fig:theta_sgl}, \ref{fig:phi_sec}, and \ref{fig:theta_sec} show the direction of travel of the photons as they escape the lamp, but not the position along the lamp periphery at which they exit. 

In the next experiment, we irradiated the lamp at a single point along the perimeter of the lamp ($\phi = 0\degr$) with both the polar and azimuthal angle kept constant. The experiment was repeated for different initial azimuthal angles from 0$\degr$ (perpendicular to the surface of the lamp) to 89$\degr$ (almost parallel to the surface). This experiment was performed using a 2 dimensional model of the lamp.

\begin{figure}[!htbp]
\centering
\subfigure[Incidence = 0$\degr$]
{\includegraphics[width=0.49\textwidth]{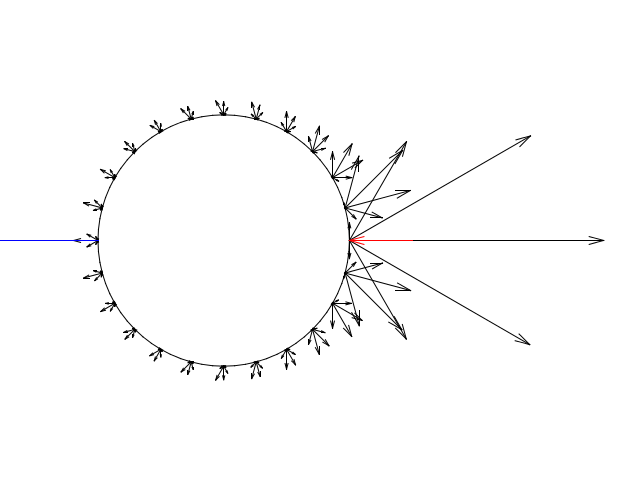}}
\subfigure[Incidence = 30$\degr$]
{\includegraphics[width=0.49\textwidth]{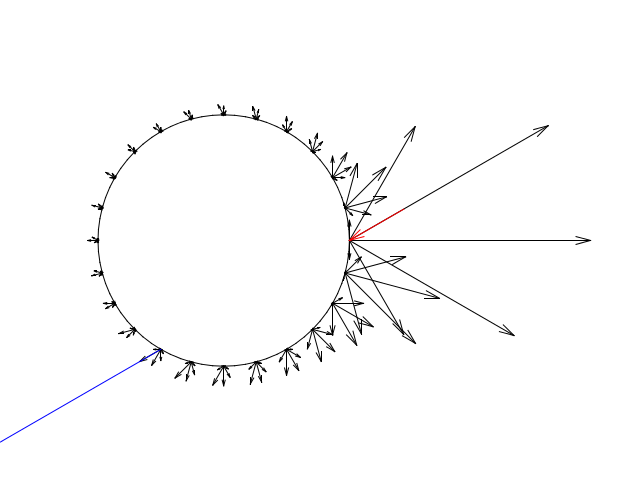}}
\\
\subfigure[Incidence = 60$\degr$]
{\includegraphics[width=0.49\textwidth]{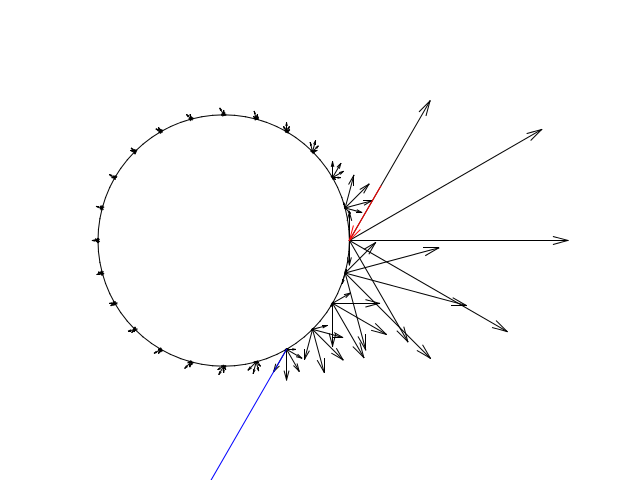}}
\subfigure[Incidence = 89$\degr$]
{\includegraphics[width=0.49\textwidth]{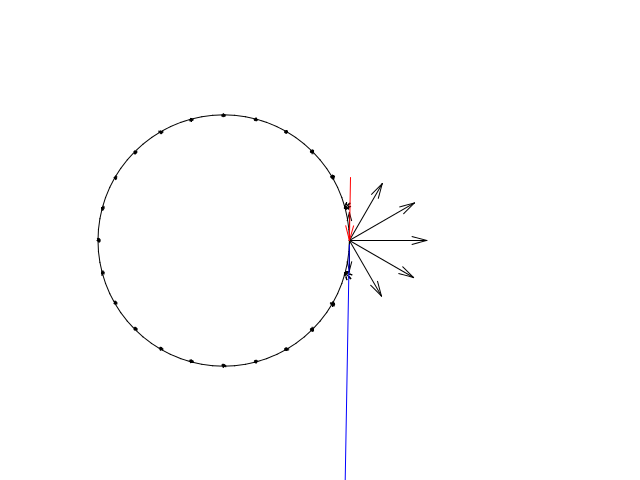}}
\caption{Intensity distribution of position and direction angles of exiting photons for incidence angles of 0$\degr$ (a), 30$\degr$ (b), 60$\degr$ (c), and 89$\degr$ (d). Black arrows represent diffused photons, while the blue arrow represents the transmitted photons and the red arrow shows where the incoming photons enter the lamp.}
\label{fig:circles}
\end{figure}

Figure~\ref{fig:circles} shows a cross-section of the lamp, with red arrows showing the direction of the incoming light, and black arrows showing the approximate position along the circumference and direction of travel of escaping photons. The relative size of the black arrows indicates the relative number of photons escaping at those positions and directions. Finally, the blue line represents the path of photons that travel through the lamp without interacting with the Hg vapour.

\subsection{Wavelength}
\label{sub:wavelength_test}

Until now, our model assumed that the vapour had a single type of Hg isotope ($^{198}$Hg) and that each of these isotopes emitted at the same wavelength, 253.6517 nm, corresponding to the central wavelength (see Table~\ref{table:isotope}). The next steps were therefore to first, increase the number of isotope species of Hg in the vapour (as listed in Table~\ref{table:isotope}), and second, broaden the emission profile of these isotopes from a single wavelength to a wavelength distribution, as discussed in Section~\ref{sub:lineshape}.

\begin{figure}[!htbp]
\centering
\includegraphics[width=0.8\textwidth]{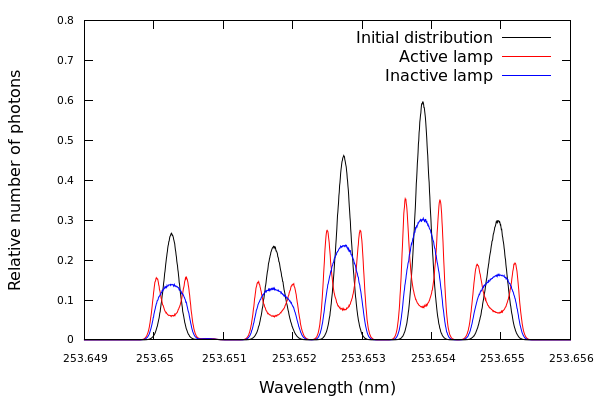}
\caption{Spectral distribution for the initial distribution, the active lamp and the inactive lamp as a function of wavelength. }
\label{fig:spectra}
\end{figure}

Figure~\ref{fig:spectra} shows three spectral distributions. The black line shows the wavelength distribution of photons as they are generated, i.e. the natural wavelength distribution. The red and blue lines show the distribution coming out of the active and inactive lamp, respectively.

\begin{figure}[H]
\centering
\includegraphics[width=0.6\textwidth]{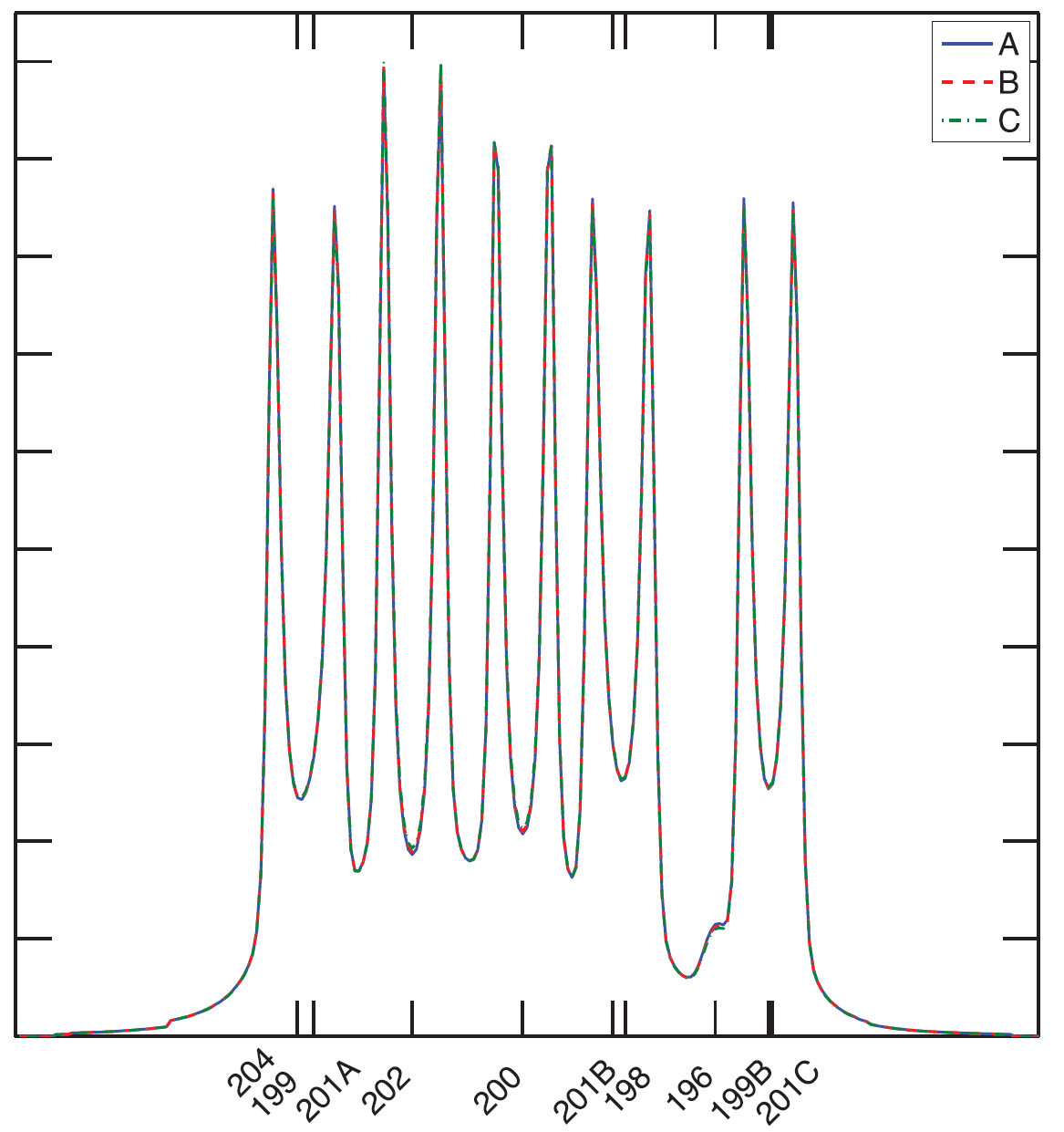}
\caption{Spectrum for different quenching rates obtained through numerical lamp modelling by Anderson (2016)\cite{and16}. The $x$-axis shows the central position of the Hg isotopes ordered by frequency, while the $y$-axis is the relative intensity. Figure reproduced without permission.  } 
\label{fig:and06}
\end{figure}

According to Equation~\ref{eq:k_dop}, the lineshape value is larger at the centre of the lines, meaning that photons near these wavelengths are more likely to be generated during the emission process, but are also more likely to be absorbed by the Hg vapour before being able to leave the lamp. On the other hand, photons farther from these central wavelengths are more likely to escape the lamp without interacting with the Hg vapour, but are also much less likely to be emitted. Because of this, and the fact that photons do not retain their wavelength after an absorption-emission event, a selection process is created, favouring photons with wavelengths that have a high enough probability of being created while having a low enough probability of being absorbed before leaving the lamp. This selection is more prevalent the longer photons stay inside the lamp. In the scenario modelled here, the majority of photons spend more time (undergoing absorption-emission events) in the active lamp than in the inactive lamp, which is why we see this effect more prominently in the active-lamp test (red line).

The so-called ``line splitting" exhibited by the active-lamp spectrum in Figure~\ref{fig:spectra} has been observed both experimentally and numerically in various other works\cite{bae03}\cite{and85}\cite{and16}\cite{zor08}\cite{raj04}. For comparison, Figure~\ref{fig:and06} shows the spectrum obtained by Anderson (2016)\cite{and16} using their own lamp models which does show splitting in the lines. 

\subsection{Longitudinal motion and wall effects}
\label{sub:long_wall}

\begin{figure}[!htbp]
\centering
\includegraphics[width=0.8\textwidth]{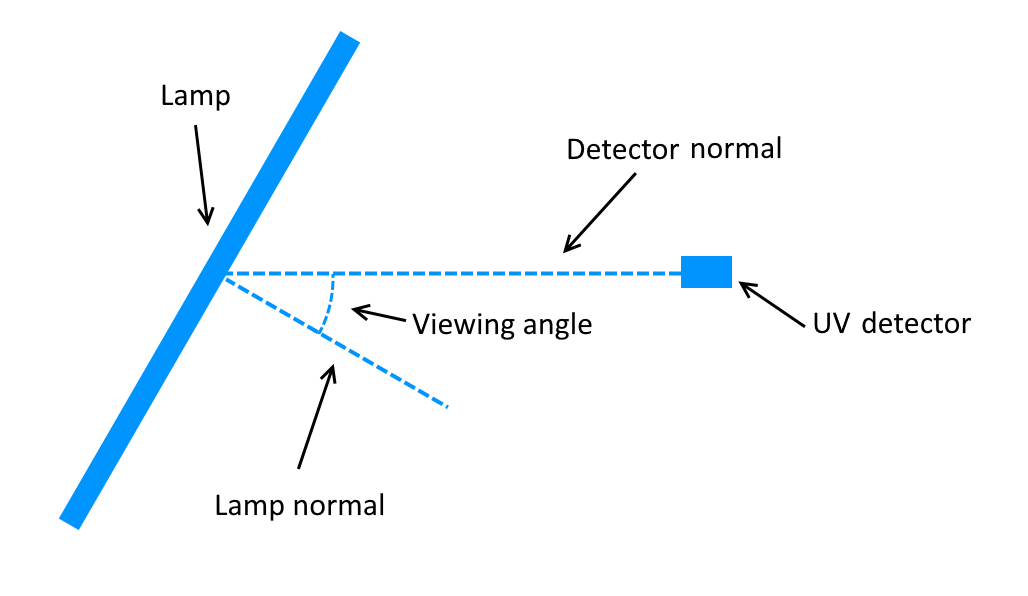}
\caption{Experimental setup for the UV irradiance measurement.}
\label{fig:lamp_schm}
\end{figure}

\begin{figure}[!htbp]
\centering
\includegraphics[width=0.9\textwidth]{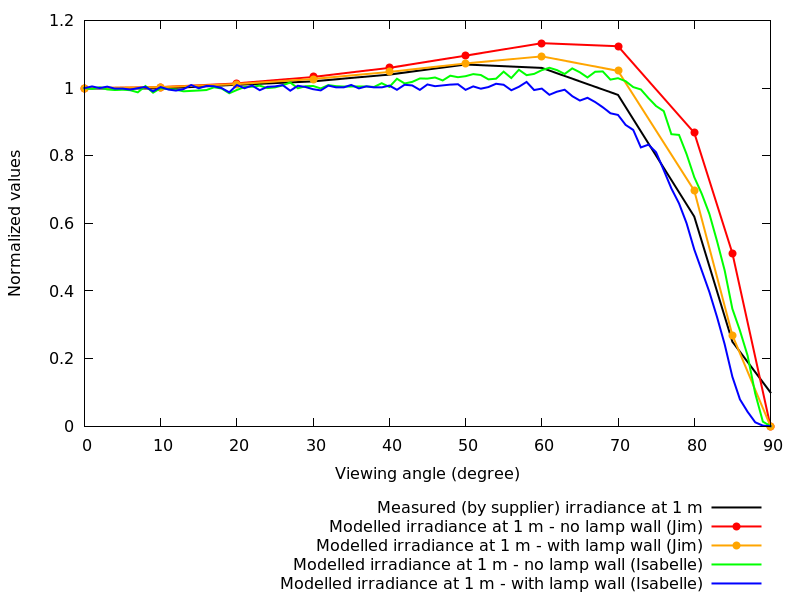}
\caption{Comparison of the relative irradiance as a function of viewing angle between experimental measurements and various models made by Jim and Isabelle for a lamp of length 1.47 m and a detector distance of 1 m.}
\label{fig:comp3d-1-2}
\end{figure}

\begin{figure}[!htbp]
\centering
\includegraphics[width=0.9\textwidth]{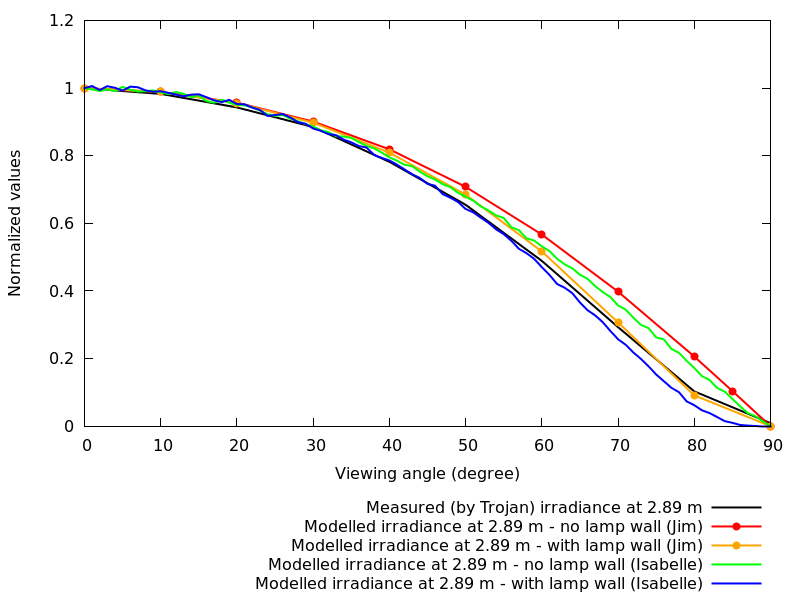}
\caption{Same as Figure~\ref{fig:comp3d-1-2} but for a detector distance of 2.89 m.}
\label{fig:comp3d-2.89-2}
\end{figure}

Finally, we added a third dimension to the tracking of photon position as well as the optical effects of the wall described in Section~\ref{sub:wall}. Figures~\ref{fig:comp3d-1-2} and \ref{fig:comp3d-2.89-2} show the relative irradiance as a function of viewing angle, as measured by a UV detector at a distance of 1 m and 2.89 m, respectively, from a 1.47 m long lamp. For this experiment, the lamp was horizontal and was rotated, in its horizontal plane, about its centre. The detector was positioned in the horizontal plane of the lamp and was aimed toward the centre of the lamp. The viewing angle was defined as indicated in Figure~\ref{fig:lamp_schm}. 

The figures show experimental results (obtained by either the lamp supplier (Figure~\ref{fig:comp3d-1-2}) or by Trojan (Figure~\ref{fig:comp3d-2.89-2})), of numerical modelling performed by Jim Robinson, and numerical results obtained from our code. We simulated measurements of irradiance using models with and without the quartz wall. We remind the reader that irradiance is the radiant flux (power) received by a surface per unit area. 

In Jim’s model, UV radiation (254 nm) is emitted from a large number of points along the lamp axis. Emission from each point is strongest in the direction normal to the lamp axis (zero angle), and falls cosinusoidally with angle from the normal. In the case in which the quartz lamp wall is included, reflection and refraction at the wall interfaces, and absorption in the quartz wall, are all taken into account.

\return

We do see bigger discrepancies between models and measurements in Figure~\ref{fig:comp3d-1-2} when the detector was closer to the lamp and the lamp measurement was provided by an external lab. Of course, some degree of discrepancy is expected because of both imperfection in the models and error in the measurements.

\subsection{Exit probability vs $^{196}$Hg concentration}

To test the validity of our models, we decided to recreate a numerical experiment performed by Anderson 1985 \cite{and85} with their own model. In this experiment, the concentration of the $^{196}$Hg isotope was varied to see how it affected the exit probability of photons, i.e. the probability that a photon generated inside the lamp has of escaping it without being quenched. For this test, we modeled a T12 type lamp (see Table~\ref{table:lamps} below for more details), complete with the quartz wall.

\begin{figure}[!htbp]
\centering
\includegraphics[width=0.8\textwidth]{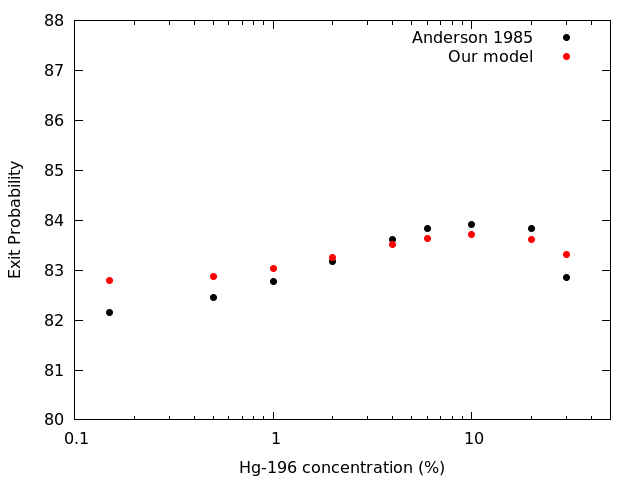}
\caption{Exit probability of photons as a function of $^{196}$Hg concentration for a T12 type lamp. The black dots show the numerical results obtained by Anderson 1985 while the red dots show the results of our numerical model. }
\label{fig:and_test}
\end{figure}

Figure~\ref{fig:and_test} compares the results of the numerical tests done by Anderson (black dots) with ours (red dots). Values from Anderson 1985 were estimated from their paper in order to be able to compare them in the same figure. 

Results from our model follow results obtained by Anderson 1985 very well, with differences in exit probabilities of less than 1\% across the board. Of course we cannot expect two different numerical models to behave the exact same way, especially when both rely on a Monte Carlo technique, which is inherently prone to small variation even between results from the same model. Furthermore, given the fact that the data from Anderson 1985 had to be extrapolated from their figure in order to be extracted, it's possible that the Anderson data in Figure~\ref{fig:and_test} might be slightly off.

\subsection{Curved reflector test}

This experiment was performed to better verify the validity of the model. The experiment consisted of placing a curved reflector, covering half the lamp’s circumference, near the longitudinal centre of the lamp. Detectors were placed at several points around the circumference of the lamp, approximately halfway along the length of the reflector, to measure the intensity at different positions relative to the reflector. The reflector surface was a curved sheet of aluminized foil with a measured reflectance at 254 nm of 93\%. The purpose of using this reflector is to give us a source of "external" photons without the complexity of adding a second lamp.

Two sets of measurements were performed; one with the reflector in place and one without it. This latter was used as a reference measurement. With these measurements, we calculated the relative increase in the measured intensity at each detector due to the presence of the reflector. 

Five experiments were performed, both in the lab and numerically, for a different set of Hg lamps and reflector length. Table~\ref{table:reflector} shows the parameters of each experiment while the properties of each lamp are presented in Table~\ref{table:lamps}.

\begin{table}[!htbp]
\begin{center}
\caption{Parameters of the curved reflector experiments}
\begin{tabular}{ lcc} 
\hline \hline \\[-2ex]
 Test & Lamp type & Reflector length (cm)\\
\hline \\[-1.8ex]

1 & T5 (UV3000) &  8.0\\

2 & T6 (UV3000Plus) &  1.5\\

3 & T6 (UV3000Plus) &  9.0\\

4 & T12 (Solo) &  7.5\\

5 & T12 (Solo) &  10.5\\

\\[-1.8ex] \hline \hline 

\end{tabular}
\label{table:reflector}
\end{center}
\end{table}

\begin{table}[!htbp]
\begin{center}
\caption{Parameters used for different lamp types}
\begin{tabular}{ccccccc} 
\hline \hline \\[-2ex]
& \multicolumn{2}{c}{Diameter} & Wall & Gas & Hg number  & Wall absorption\\
Lamp type & Inner & Outer & Thickness & Temperature  & density & coefficient \\
& (mm) & (mm) & (mm) &($\degr$C) & (10$^{14}$ atoms/cm$^3$) & (base 10, cm$^{-1}$)\\
\hline \\[-1.8ex]

T5  &  - & 16 & -  & 72 & 3.2 & 0.897\\
(UV3000) & & & & & &\\
T6  & -  & 19 & -  & 125 & 2.1 & 0.175\\
(UV3000Plus) & & & & & &\\
T12 &  - & 38 & -  & 215 & 1.3 & 0.175\\
(Solo) & & & & & &\\

\\[-1.8ex] \hline \hline  	

\end{tabular}
\label{table:lamps}
\end{center}
\end{table}

\begin{figure}[!htbp]
\centering
\includegraphics[width=0.8\textwidth]{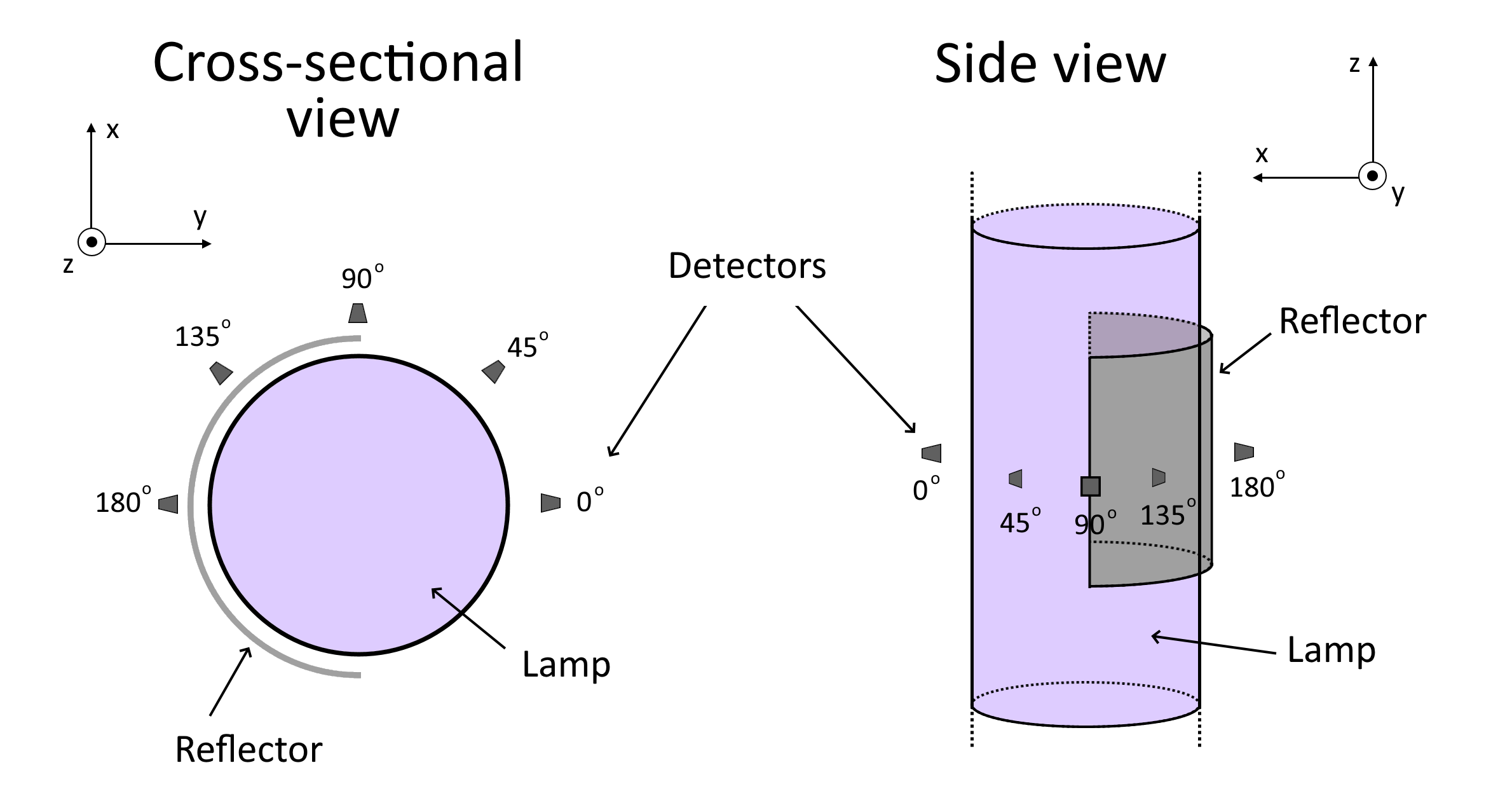}
\caption{Diagram of the experimental setup for the curved reflector test.}
\label{fig:reflector}
\end{figure}

Figure~\ref{fig:reflector} shows a schematic of the experiment seen from a cross-sectional view and a side view. 

The experimental setup used optical fibres to carry light to the detectors. For measurements at 180$\degr$, a small hole was made in the reflector to let the fibre through. However this was not done for all experiments, therefore only 3 out of the 5 tests have measurements for 180$\degr$. We also note that no experimental measurements were performed for detectors at 45$\degr$ or 135$\degr$.
 
\begin{figure}[!htbp]
\centering
\subfigure[Test 1]
{\includegraphics[width=0.48\textwidth]{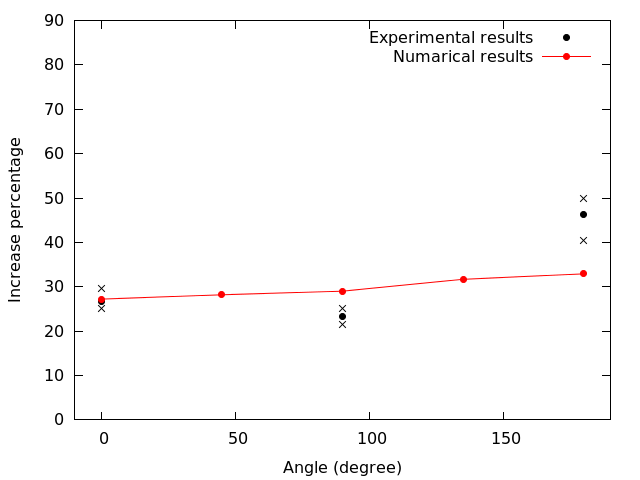}}\\
\subfigure[Test 2]
{\includegraphics[width=0.48\textwidth]{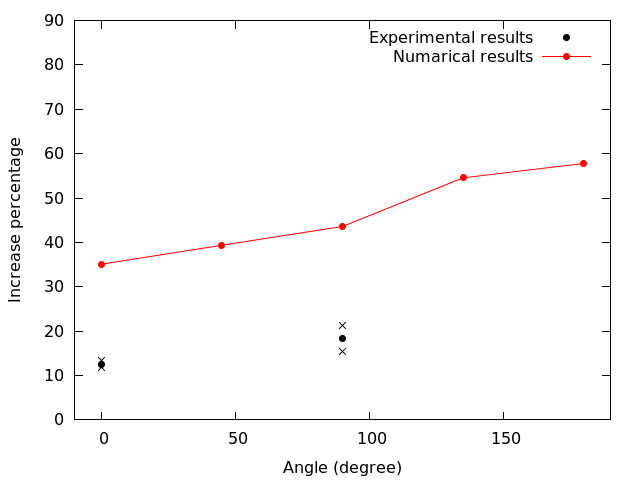}}
\subfigure[Test 3]
{\includegraphics[width=0.48\textwidth]{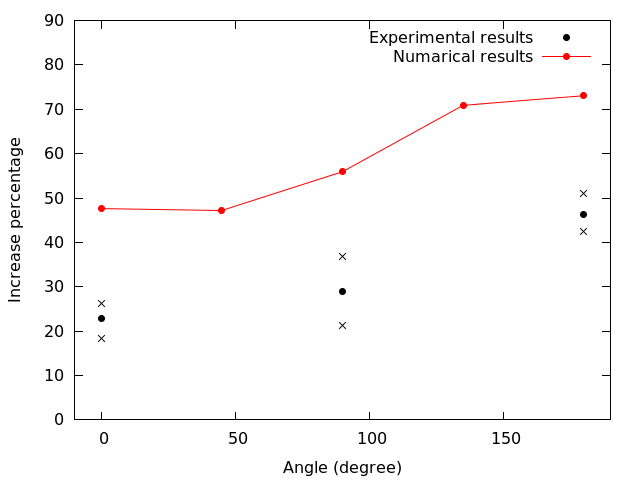}}\\
\subfigure[Test 4]
{\includegraphics[width=0.48\textwidth]{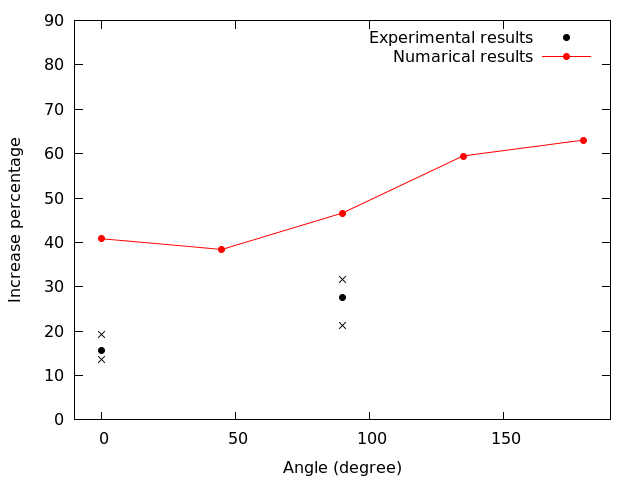}}
\subfigure[Test 5]
{\includegraphics[width=0.48\textwidth]{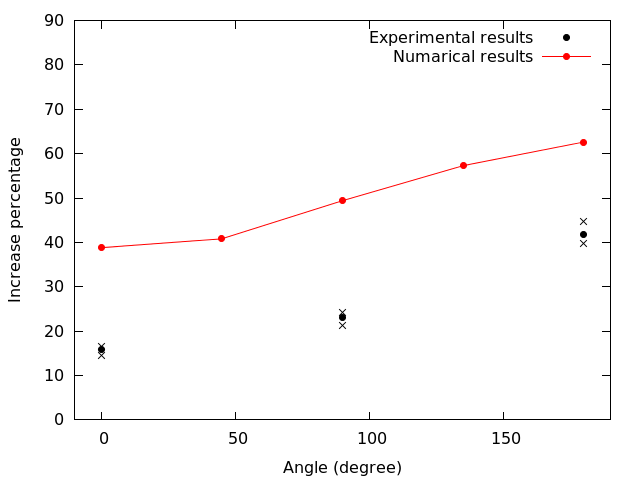}}
\caption{Relative increase in intensity as a function of detector position for the curved reflector test. Each panel shows the results of a different test as described in Table~\ref{table:reflector}. The black dots and crosses are the means and ranges, respectively, of the laboratory experiments, the red lined dots show the results of the numerical simulations. }
\label{fig:ref_results}
\end{figure}

Figure~\ref{fig:ref_results} shows a comparison between the experimental and numerical results of the curved reflector test. 

Overall the numerical results show a good agreement with the experimental results. Deviations between results are more pronounced for the detector at 180$\degr$ but this is to be expected since this detector had to be moved between tests in order to place/remove the reflector, leading to different position or orientation of the fibre between measurements. It is important to note that some of the parameters of the lamps were estimations, which can also lead to divergence between results. 

\subsection{Two-lamp test}

\begin{figure}[!htbp]
\centering
\includegraphics[width=0.8\textwidth]{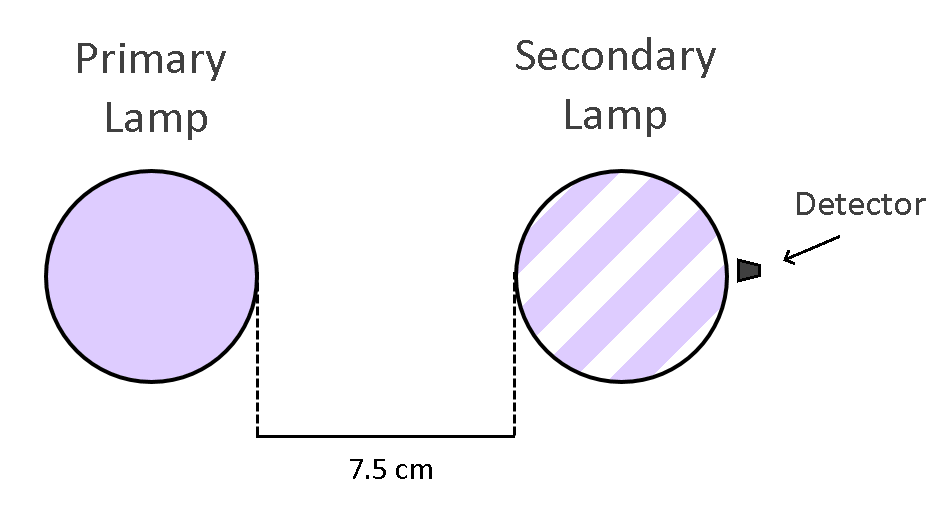}
\caption{Diagram of the experimental setup for the two-lamp test.}
\label{fig:2lamps}
\end{figure}

The final test we performed consisted of placing two lamps parallel to each other, approximately 7.5 cm apart. The primary lamp (T6 lamp) served as the source, irradiating the secondary one (T5 lamp). A detector was placed at the surface of the secondary lamp, on the opposite side of the primary, so that it pointed towards both lamps. Figure~\ref{fig:2lamps} is a graphical representation of this setup. 

This experiment was performed for three different configurations. In the first, the secondary lamp was an inactive Hg filled lamp. In the second configuration, the secondary was changed to a lamp of the same dimension but containing no Hg inside. For the final configuration the secondary lamp was removed altogether, leaving only the primary. For all three configurations the primary and the detector remained unchanged. 

To help compare the experimental results with the numerical ones, ratios of the measurements were taken. These ratios are filled lamp to empty lamp (Hg/Non-Hg) and filled lamp to no lamp (Hg/None). Results for these ratios are presented on Table~\ref{table:2lamps_res}. Once again we note that the results given by our numerical model agree well with the laboratory measurements.

\begin{table}[!htbp]
\begin{center}
\caption{Experimental and numerical results of the two-lamp test}
\begin{tabular}{ lcc} 
\hline \hline \\[-2ex]
 & Experimental & Numerical\\
 Ratio & result & result\\

\hline \\[-1.8ex]

Hg/Non-Hg &   0.43 & 0.40\\

Hg/None &   0.24 & 0.26\\

\\[-1.8ex] \hline \hline  	

\end{tabular}
\label{table:2lamps_res}
\end{center}
\end{table}

\section{Conclusions}
\label{summary}

We have developed a Monte Carlo code that can follow the interactions of UV radiation with neighbouring lamps and can predict the amount of UV that is quenched and re-emitted. The code is flexible to allow different sizes of lamps, orientations and distances between lamps to be investigated. Overall we find that our lamp model is robust enough to replicate certain experimental and numerical results found in the literature. 

\section{Future Plans}
\label{future}

The final step remaining for this project is to implement an algorithm, using the modeling tools we have developed, to quickly compute the resulting output rays from a single incident ray. The output will be formatted so that it is easily readable by the ray tracing code \textit{TracePro}. This step is already under way and should be finished by the end of January 2021.

\bibliography{ref.bib} 

\begin{thebibliography}{10}

\bibitem{mol98}
A.~Molisch and B.~Oehry, {\em Radiation Trapping in Atomic Vapours}.
\newblock Oxford science publications, Clarendon Press, 1998.

\bibitem{ber69}
P.~R. Berman and W.~E. Lamb, ``Influence of resonant and foreign gas collisions
  on line shapes,'' {\em Phys. Rev.}, vol.~187, pp.~221--266, Nov 1969.

\bibitem{and85}
J.~Anderson, J.~Maya, M.~W.~Grossman, R.~Lagushenko, and J.~F.~Waymouth,
  ``Monte carlo treatment of resonance-radiation imprisonment in fluorescent
  lamps,'' {\em Physical review. A}, vol.~31, p.~2968 to 2975, Jun 1985.

\bibitem{mit61}
A.~Mitchell, A.~Mitchell, and M.~Zemansky, {\em Resonance Radiation and Excited
  Atoms}.
\newblock Cambridge series of physical chemistry, Cambridge University Press,
  1961.

\bibitem{bae03}
M.~Baeva and D.~Reiter, ``Monte carlo simulation of radiation trapping in
  hg--ar fluorescent discharge lamps,'' {\em Plasma Chemistry and Plasma
  Processing}, vol.~23, p.~371 to 387, Jun 2003.

\bibitem{vei83}
J.~Veizer, ``Trace elements and isotopes in sedimentary carbonates,'' {\em
  Reviews in Mineralogy}, vol.~11, Feb 1983.

\bibitem{hay12}
W.~Haynes, {\em CRC Handbook of Chemistry and Physics, 93rd Edition}.
\newblock 100 Key Points, Taylor \& Francis, 2012.

\bibitem{hub06}
M.~L. Huber, A.~Laesecke, and D.~G. Friend, ``Correlation for the vapor
  pressure of mercury,'' {\em Industrial \& Engineering Chemistry Research},
  vol.~45, no.~21, p.~7351 to 7361, 2006.

\bibitem{box58}
G.~E.~P. Box and M.~E. Muller, ``A note on the generation of random normal
  deviates,'' {\em Ann. Math. Statist.}, vol.~29, pp.~610--611, 06 1958.

\bibitem{lee74}
J.-S. {Lee}, ``Monte carlo simulation of voigt distribution in photon diffusion
  problems,'' {\em The Astrophysical Journal}, vol.~187, pp.~159--162, Jan.
  1974.

\bibitem{hec02}
E.~Hecht, {\em Optics}.
\newblock Pearson education, Addison-Wesley, 2002.

\bibitem{mik12}
A.~Mik\v{s} and P.~Nov\'{a}k, ``Determination of unit normal vectors of
  aspherical surfaces given unit directional vectors of incoming and outgoing
  rays: comment,'' {\em J. Opt. Soc. Am. A}, vol.~29, pp.~1356--1357, Jul 2012.

\bibitem{and16}
J.~B. Anderson, ``Monte carlo treatment of resonance-radiation imprisonment in
  fluorescent lamps{\textemdash}revisited,'' {\em Journal of Physics D: Applied
  Physics}, vol.~49, p.~495501, Nov 2016.

\bibitem{zor08}
N.~Zorina, G.~Revalde, and R.~Disch, ``Deconvolution of the mercury 253.7 nm
  spectral line shape for the use in absorption spectroscopy,'' {\em
  Proceedings of SPIE - The International Society for Optical Engineering},
  vol.~7142, Nov 2008.

\bibitem{raj04}
K.~Rajaraman and M.~J. Kushner, ``A monte carlo simulation of radiation
  trapping in electrodeless gas discharge lamps,'' {\em Journal of Physics D:
  Applied Physics}, vol.~37, p.~1780 to 1791, Jun 2004.

\end{thebibliography}
\bibliographystyle{ieeetr}

\end{document}